# Thermodynamic model of interaction of small ligands with DNA.

Vasil G. Bregadze*, Eteri S. Gelagutashvili, Irene G. Khutsishvili, Khatuna G. Sologashvili, Ketevan J. Tsakadze.

Andronikashvili Institute of Physics, 6 Tamarashvili st., Tbilisi, Georgia 0177;

∗Address for correspondence: e-mail: **breg@iphac.ge**

**v_breg@yahoo.com**

Abstract

We have managed to correlate stability constants for complex formation K, which can be registered in equilibrium state, to dynamic characteristic $\tau$ for the lifetime of a complex $\tau = \tau_0 \cdot K$. Thus, the principal concept of molecular biophysics regarding biomolecule: structure-dynamics-function can be reformatted as structure-stability-function. It should be specially noted that such an approach highly simplifies end widens time interval (from $10^{-10}$ to $10^5 - 10^6$ sec. and more) under investigation of dynamic characteristics of macromolecules.

Study of hydration energy and hydration number by kinetic curves of water glow discharge atomic spectral analysis of hydrogen (GDAS analysis of hydrogen) desorption from Na-DNA humidified fibers allowed to reveal that structural and conformational changes in activation energy of hydrated water molecules increases by 0.65kcal/Mole of water. This increase of energy is 0.1 kcal/Mole for B→C transition (from 20 to 10 water molecules per nucleotide) and 0.55 kcal/Mole for C→A unordered state (from 10 to 5 molecules per nucleotide).

Simultaneous use of UDS and AES allows identifying number of WWC (Wrong Watson-Crick) pairs in DNA. The quantitative analysis is based on increase of informativity of UDS via its combination with atomic-emission spectroscopy. The number of the total metal-induced tautomerization makes ~585 per $10^6$ base pairs of DNA. Such approach enables estimation of risk factors during clinical check-ups and evaluation of DNA suitability for nanotechnological purposes.

## Introduction

There are number of small ligands that change DNA stability [1,2], cause certain defects (depurinsation [3], double proton transfer in G-C pairs [2,4]) and also participate in various processes such as energy, electron and a proton transfer, in which DNA molecule plays a role of a mediator *in vitro* [5 – 13].

Especially interesting from this point of view is macro- and micro-elemental composition of a real DNA molecule. There were many works [14-17] devoted to this problem. It was noted tissue specificity and dependence on carcinogenicity of DNA [18].

New original thermodynamic model of interaction of small ligands with DNA establishes direct proportional correlation between the dynamical characteristics of interaction, i.e. the lifetime of the complex, and its equilibrium characteristic – the stability constant. We have obtained simple expression for description of the time for small ligand interaction with macromolecules

$$\tau = \tau_0 \cdot K , \qquad (1)$$

where $\tau$ is lifetime of a ligand - macromolecule complex [2]. The parameter $\tau_0$ is analogous to Frenkel's $\tau_0^{a-s}$, which is duration of the fluctuation excitation of the adsorbing atoms or molecules interacting with a solid surface, and it is supposed to be equal to the period of the oscillation of the adsorbate relative to the adsorbent surface. In solutions, the value $\tau_0$ describes duration of the relaxation of rotary and translation movements of the solvent molecules, ions, solvated ions or low-molecular-weight substances and lies between $10^{-11}$ and $10^{-10}$ sec. (Tab. 1).

Tab. 1. Characteristics of Inner Movements in DNA

| Type of movement | Time, sec | Excitation energy, *kcal/mol* |
|---|---|---|
| 1. Various small – amplitude oscillating movements of atoms (about 0,1 $\overset{o}{A}$) inside the components of DNA | $10^{-14}$—$10^{-13}$ ($3500$-$300 cm^{-1}$) | RT(T=300°K)=0.6(210 $cm^{-1}$) |
| 2. Limited movements of phosphates, sugars and nucleobases relative to the equilibrium position, torsional and flexural oscillations of the double helix) | $10^{-10}$—$10^{-8}$ | |
| 3. Large –amplitude movements of phosphates, sugars and bases occurring in connection with the transition of the double helix from one form to another | $10^{-7}$—$10^{-5}$ | 5-6 for B$\rightarrow$A transitions and 21 for B$\rightarrow$Z transitions [19] |
| 4. Change of free energy $\Delta$G needed for opening of central pairs in double chain RNA at 25°C, kcal/mol for bases [20] | | |
| GC $\begin{pmatrix} G—G—G \\ \mid \quad \mid \quad \mid \\ C-C-C \end{pmatrix}$ and | ~3*$10^{-6}$ *) | 7.5 |
| AU $\begin{pmatrix} A—A—A \\ \mid \quad \mid \quad \mid \\ U-U-U \end{pmatrix}$ | ~$10^{-8}$ *) | 4.0 |

*) The time of pair opening is evaluated by formulae (1)

Thus, if we assume *logK=4-6* for binding of DNA with twofold positively charged metal ions of the first transition series, and $\tau_0 = 10^{-11}$ sec, then the life span of these complexes is about $10^{-7}$ – $10^{-5}$ sec.

Therefore, we have managed to correlate stability constant for complex formation *K*, which can be registered in equilibrium state, to dynamic characteristic $\tau$ for the lifetime of a complex. Thus, a principal concept of molecular biophysics **structure – dynamics – function** regarding biomolecule, can be reformulated as a concept of **structure – stability – function**. It should be specially noted that such approach highly simplifies and widens the time interval (from $10^{-10}$ sec to $10^5$ – $10^6$ sec and more) of assessable dynamic characteristics of macromolecules under investigation.

**The Investigation of Energetics of Water Molecule Binding with Biopolimers.** One of the interesting applications of the VHF inductively coupled plasma of reduced pressure is its use as a light source in atomic-emission kinetic analysis of water vapor desorbed from the humidified samples of biological origin in order to study the kinetics and energetics of the water molecules absorbed by their surface [21]. In this case, the spectrophotometer tuned to one of the lines of Balmer series $H_\alpha$ = 656.284 nm, or $H_\beta$ = 486.133 nm of hydrogen atom records the quantity of water vapours passing through the VHF-plasmatron discharge tube over the time. This is a new method of investigation of biopolymer hydration. Below, on the example of kinetics of water desorption from DNA, we shall consider two modifications of this method – isothermal and flash-desorption.

**Double Proton Transfer.** UV spectroscopic display of self-congruent double proton transfer (DPT) in GC pairs of thymus DNA induced by H+ and transition metals (TM) was studied[13].The choice of GC pair was proved by the following:

1. Guanine and cytosine are more populated with rare enolic- and imino- forms [22-24] and these forms are easily observed in UV range of spectrum;

2. GC pairs are far less resistant to tunneling transitions compared with AT pairs [4,25];

3. Endocyclic nitrogen of pyridine type $N_7$ of guanine located in the major groove of DNA duplex is the preferable binding site for $H^+$ and TM [13].

UV spectroscopic display of double proton transfer (DPT) in GC pairs of DNA induced by local effect of $H^+$ ions and transition metal ions as well as by macroscopic change of its surroundings (high concentration of salt and ethanol, polyethylenglycol, etc. was studied in [13]. We have used phenomenological quantum – mechanical approach to explain DPT mechanism as the effect of the environment on electron configuration of atoms in molecules participating in H – bonds. This consideration takes into account specific energy profile of double – level H – bonds, and self – congruent transition of protons in cyclic structures (Fig.1):

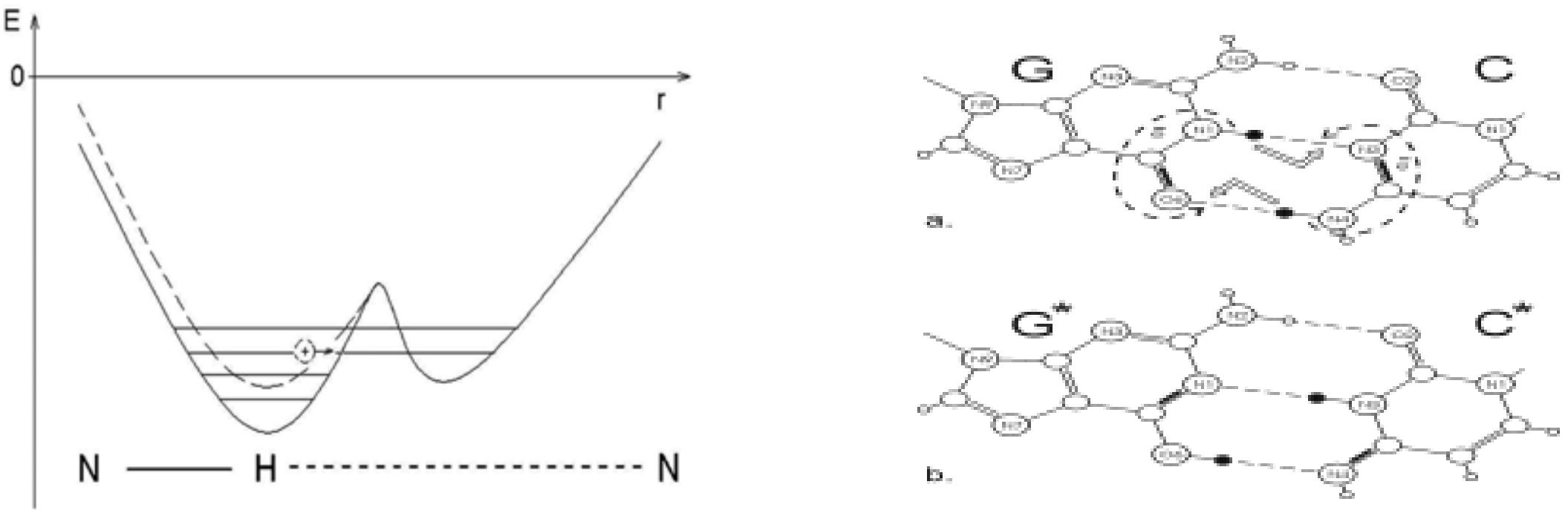

A. B.

Fig. 1. Double proton transfer in DNA G-C pair.

A. The hypothetic function of potential energy of N- H ···N type bond;

B. Self-congruent transition in Watson-Crick G-C base pair leading to formation of wrong $G^*$-$C^*$ pair.

These structures form unified electronic – proton complementary molecular complex between proton – donor – acceptor groups. The last is necessary and satisfactory condition for keto – enolic and amino – imino tautomeric transformations in solutions of polar organic molecules.

We have investigated energetic characteristics of interactions of different metal ions with DNA. In addition, in order to reveal the influence of ionic strength and G-C composition on this parameter, bonding constants for $Cu^{2+}$ with DNA of different origin, histone core, and different ionic strengths were determined. The Table 1 represents the results of this study.

Tab. 2. Influence of ionic strength, histone core and G-C composition on bonding constants for $Cu^{2+}$ with DNA of different origin.

| DNA | Ion | Ionic strength, mM | pK |
|---|---|---|---|
| Calf thymus DNA (40% G-C) | $Co^{2+}$ | 20 | 4.30 |
| “ ---------------------------“ | $Ni^{2+}$ | 20 | 4.64 |
| “----------------------------“ | $Zn^{2+}$ | 20 | 4.20 |
| “ ---------------------------“ | $Cu^{2+}$ | 20 | 5.10 |
| “ ---------------------------“ | $Cu^{2+}$ | 2 | 5.40 |
| “ --------------------------“ | $Cu^{2+}$ | 200 | 4.6 |
| Calf thymus DNA (40% G-C) | $H^+$ | 10 | 3.7 |
| DNA from C3HA mice liver(40% G-C) | $Cu^{2+}$ | 10 | 4.98 |
| DNA from ascetic hepatoma 22A of C3HA mice(40% G-C) | $Cu^{2+}$ | 10 | 5.07 |
| Mononucleosomes from C3HA mice liver(40% G-C) | $Cu^{2+}$ | 10 | 4.15 |
| Mononucleosomes from ascetic hepatoma of 22A of C3HA mice (40% G-C) | $Cu^{2+}$ | 10 | 4.24 |
| Cancerous human breast tumor (40% G-C) | $Cu^{2+}$ | 20 | 5.30 |
| Non-cancerous human breast tumor(40% G-C) | $Cu^{2+}$ | 20 | 4.91 |
| DNA from bacteriophage T4 (36% G-C) | $Cu^{2+}$ | 20 | 4.86 |
| DNA from clostridium perfringens (27% G-C) | $Cu^{2+}$ | 20 | 4.01 |

Bonding constants (*logK*) for $Mg^{2+}$, $Mn^{2+}$, $Co^{2+}$, $Ni^{2+}$, $Cu^{2+}$ and $Zn^{2+}$ ions with DNA change from 4 to 6, and if $\tau_0$ is considered to be $10^{-11}$ sec then $\tau$ lifetime, according to formulae (1) will be $10^{-7}$ – $10^{-5}$ sec. For DNA, this time corresponds to the movements of phosphates, sugars and bases with big amplitudes (Tab. 3), these movements are connected to the change of the form of duplex, untwisting of double helix. As metal ions of first transition series form chelate complexes with DNA the stability of the complexes is estimated by DNA dynamics. Even more, macroscopic thermodynamic stability constants [2, 26] of $Co^{2+}$, $Ni^{2+}$, $Cu^{2+}$ and $Zn^{2+}$ ions with thymus DNA are equal to 5.21, 5.52, 5.80 and 5.15 correspondingly.

It means that the difference between $Co^{2+}$ and $Cu^{2+}$ makes only 0.59 logarithmic units when for the interaction of the said ions with bidental ligand $NH_2CH_2CO_2$ it varies from 4.6 to 8.6 and in the case of ethylene diamine – from 6 to 11 [27]. Thus when $M^{2+}$ ions interact with DNA, the dynamic properties of DNA connected with the big amplitude for the unwinding of the double helix and the opening of the base pairs determine both the values of the stability constants of the DNA complexes with $Co^{2+}$, $Ni^{2+}$ , $Cu^{2+}$, and $Zn^{2+}$ and the limits of their changes.

Usually, a serious problem exists in estimation of bonding constants of DNA complexes with high stability constants (*logK*>8-10). These types of ions include high toxic for the living organisms $Ag^+$ and $Hg^{2+}$. They have neither EPR signal nor absorption spectra in near UV and visible ranges of spectrum. Developed by us a new method of the joint use of kinetic and equilibrium approaches to kinetics of $Cu^+$ ion oxidation in complexes with DNA and comparative study of UV difference spectra of DNA complexes with $Cu^+$ and $Ag^+$ allows us to estimate energy and lifetime of these complexes. The results are presented in Table 3.

Tab. 3.Energetic, equilibrium and spectral characteristics of DNA interaction with $Cu^{+}$ and $Ag^{+}$

| | E (*kcal/M*) | LogK | τ (sec.) | Δυ |
|---|---|---|---|---|
| $Cu^{+}$ | 20.5 | 14.9 | $8.6 \cdot 10^{3}$ | 700 |
| | 22.9 | 16.7 | $4.8 \cdot 10^{5}$ | |
| $Ag^{+}$ | ≥14.9 | ≥10.8 | ≥0.63 | 500 |

The model applies physical model of gas and vapor adsorption by the surface of solid bodies. It was successfully tested for estimation of dynamic characteristics of DNA related to big amplitude untwisting of double helix and lifetime of DNA complexes with $M^{2+}$ ions.

**Wrong Watson-Crick Pairs.** The self-congruent DPT does not change the geometry of the G-C pair and it may be overlooked by the reparation system causing the transition type point mutation, which may occur very dangerous for the living cells (Fig.2).

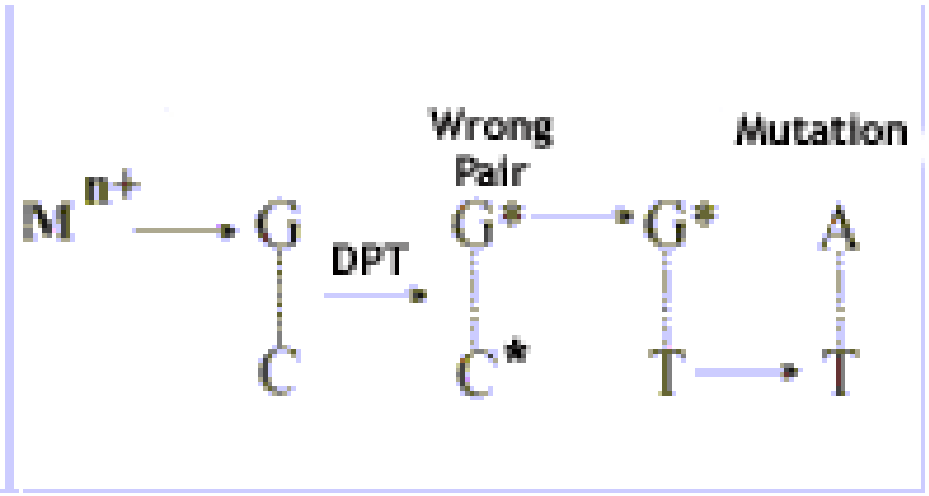

Fig.2 Schematic display of point mutation caused by DPT.

Double Proton Transfer (DPT) basically reveals itself in UF spectroscopy by batochromic shift of the absorption spectrum, slight hipochromic effect and little widening of the absorption band of DNA (Tab. 1).

Tab. 4. spectral characteristics of interaction of DNA with metal ions.

| Ions | Total extinction $\Delta\varepsilon_S^{1)}$ | $\Delta\nu^{1)}$ ($cm^{-1}$) (calculated) | $\Delta\nu^{*}$ ($cm^{-1}$) (observed) | Change of intensity | Widening ΔW |
|---|---|---|---|---|---|
| $H_3O^{+}$ | 440 | - | 300 | 2.5(decrease) | |
| $Mg^{2+}$ | 25 | 15 | - | | |
| $Ca^{2+}$ | 10 | 6 | - | | |
| $Cr^{2+}$ | 250 | 165 | - | | |
| $Mn^{2+}$ | 100 | 65 | - | | |
| $Fe^{3+}$ | 340 | 230 | - | | Slight aggregation |
| $Fe^{2+}$ | 140 | 95 | - | | |
| $Co^{2+}$ | 170 | 110 | - | | |
| $Ni^{2+}$ | 220 | 150 | - | 1.5(decrease) | |
| $Cu^{2+}$ | 290 | 190 | - | | 140 (1/2 h) |
| $Cu^{+}$ | 2400 | - | 700 | 7.4(decrease) | |
| $Zn^{2+}$ | 130 | 85 | - | | |
| $Ag^{+}$ | 1480 | - | 500 | 10(decrease) | 1000 (2/3 h) |
| $Cd^{2+}$ | 300 | 200 | - | | |
| $Pb^{2+}$ | 320 | 220 | - | | |

[a] Values of $\Delta\varepsilon_s$ and $\Delta\nu$ are given as 1 ion per 1 GC pair in Thymus DNA (40% GC), $\Delta\nu$ is calculated from expression $\Delta\nu = \Delta\varepsilon_s / K$ where $\Delta\varepsilon_s = |\Delta\varepsilon(\nu_{max})| + |\Delta\varepsilon(\nu_{min})|$ , $K = |\varepsilon^{(1)}(\nu_{max})| + |\varepsilon^{(1)}(\nu_{min})|$ at $10^{-2}$ NaCl [13 ].

[1] $\Delta\varepsilon_S$ and $\Delta\nu$ are taken with the stoichiometry 1 ion per 1 G-C pair for thymus GNA (40% G-C) ; Evaluation precision of $\Delta\varepsilon$ was $\delta_{\Delta\varepsilon}= \pm3$ for $H_3O^+$, $Cr^{2+}$, $Mn^{2+}$, $Co^{2+}$, $Ni^{2+}$, $Cu^{2+}$, $Zn^{2+}$, $Cd^{2+}$, $Pb^{2+}$; $\delta_{\Delta\varepsilon}= \pm2$ for $Ca^{2+}$; $\delta_{\Delta\varepsilon}= \pm15$ for $Ag^+$ and $\delta_{\Delta\varepsilon}= \pm15\text{-}20$ for $Cu^+$.

Table 5 represents quantitative characteristics of tautomery induced by ions in DNA and the summary spectral shift caused by protonation of guanine and cytosine. Assuming that the ions interact just with the G-C pairs, we may conclude that the number of wrong WC base pairs (n) is proportional to the ion concentration. It is also known that the long-wave shift of DNA absorption spectra by 700cm$^{-1}$ corresponds to 1M wrong G-C pairs [28]. Considering this, we may calculate n per $10^6$ base pairs of DNA using expression:

$$n= \frac{\Delta\nu_{M^{n+}}}{700} N . \qquad (2)$$

N is number of metal atoms per $10^6$ DNA bp; number of wrong Watson-Crick GC pairs per $10^6$ DNA bp.

Tab. 5. Quantitative characteristics of metal induced tautiomeryin calf thymus DNA

| Metal ions | N 2) | n 3) |
|---|---|---|
| $Mg^{2+}$ | 3500 | 75 |
| $Ca^{2+}$ | 5000 | 43 |
| $Cr^{3+}$ | 320 | 75 |
| $Fe^{3+}$ | 710 | 233 |
| $Fe^{2+}$ | 710 | 96 |
| $Co^{2+}$ | 40 | 6 |
| $Ni^{2+}$ | 230 | 49 |
| $Cu^{2+}$ | 230 | 62 |
| $Cu^{+}$ | 230 | 230 |
| $Zn^{2+}$ | 210 | 26 |
| $Cd^{2+}$ | 9 | 3 |
| $Pb^{2+}$ | 40 | 13 |

2) N is the number of microelements on $10^6$ b.p.
3) n –number wrong Watson-Crick pairs on $10^6$ b.p

In natural conditions, DNA is not surrounded by this high amount of metals. At little concentrations, DPP in fact has no effect on UVS of DNA. Nevertheless, it is highly expected that at little concentrations metal ions more easily get to $N_7$ Guanines of DNA causing tautomerization, and, consequently, mutations. From this point of view, a number of wrong GC pairs caused by endogenous metal ions, i.e. point mutation risk factor is of an interest to estimate. In the first approximation, we can assume a number of defects caused by metal ions to be proportional to ion concentration taking into account the fact that 0,04M of metals. The total number of wrong GC pairs caused by the presented in the table metals makes approximately 585 per $10^6$ pairs of DNA. In this sum defects caused by $Fe^{2+}$ and $Cu^{+}$ are not accounted as endogenously they exist only in $Fe^{3+}$ and $Cu^{2+}$ reduced forms.

R E F E R EN C E S